\documentclass[sigconf,natbib=true]{acmart}

\usepackage{multirow} 
\usepackage{xcolor}
\usepackage{tabularx}
\usepackage{balance}

\usepackage{booktabs}
\usepackage{pifont}
\newcommand{\cmark}{\ding{51}}

\newcommand{\boldheading}[1]{%
  \par\bigskip%
  \noindent%
  \textbf{#1}%
}

\definecolor{darkblue}{HTML}{4A4ABC}
\definecolor{darkred}{HTML}{C44358}
\definecolor{darkyellow}{HTML}{DBB975}
\definecolor{darkgreen}{HTML}{509359}
\definecolor{lightgreen}{HTML}{91b7a1}

\newcommand{\established}{\textcolor{green}{$\bullet$}}
\newcommand{\sota}{\textcolor{blue}{$\bullet$}}

\usepackage{tcolorbox}

\AtBeginDocument{%
  }

\copyrightyear{2026}
\acmYear{2026}
\setcopyright{cc}
\setcctype{by}
\acmConference[SIGIR '26]{Proceedings of the 49th International ACM SIGIR Conference on Research and Development in Information Retrieval}{July 20--24, 2026}{Melbourne, VIC, Australia}
\acmBooktitle{Proceedings of the 49th International ACM SIGIR Conference on Research and Development in Information Retrieval (SIGIR '26), July 20--24, 2026, Melbourne, VIC, Australia}
\acmDOI{10.1145/3805712.3808573}
\acmISBN{979-8-4007-2599-9/2026/07}

\begin{document}

\title{A Standardized Re-evaluation of Conversational Recommender Systems on the ReDial Dataset} 

 \author{Ivica Kostric}
 \affiliation{%
   \institution{University of Stavanger}
   \city{Stavanger}
   \country{Norway}
 }
 \email{ivica.kostric@uis.no}

 \author{Krisztian Balog}
 \affiliation{%
   \institution{University of Stavanger}
   \city{Stavanger}
   \country{Norway}
 }
 \email{krisztian.balog@uis.no}

\begin{abstract}
Recent years have seen a surge of research into conversational recommender systems (CRS). Among existing datasets, ReDial is the most widely used benchmark, cited in hundreds of studies. However, variations in how the dataset is preprocessed and used in experiments, particularly in the definition of ground-truth items, make it difficult to compare results across studies. These comparisons are further complicated by confounding factors such as the choice of the underlying large language model (LLM) and the use of external data sources. In this work, we revisit seven prominent CRS methods across three architectural families and evaluate them under standardized conditions. Our reproducibility study reveals a ``granularity gap,'' where fine-grained ranking (Recall@1) is highly sensitive to implementation details, while our replicability analysis shows that nearly 50\% of reported accuracy stems from ``repetition shortcuts'' that are absent in novelty-focused evaluation. Furthermore, we find that performance gains are often driven more by the capacity of the LLM backbone than by specific architectural innovations. Finally, by applying user-centric utility metrics, we demonstrate that traditional recall frequently overstates a system's actual conversational effectiveness. This work establishes a transparent, controlled baseline and promotes evaluation practices that prioritize novelty and interaction efficiency.
\end{abstract}

\ccsdesc[500]{Information systems~Recommender systems}

\keywords{Conversational Recommender Systems; Recommendation; Reproducibility; Large Language Models; ReDial}

\maketitle

\section{Introduction}
\label{sec:intro}

\begin{figure}
    \centering
    \includegraphics[width=\linewidth]{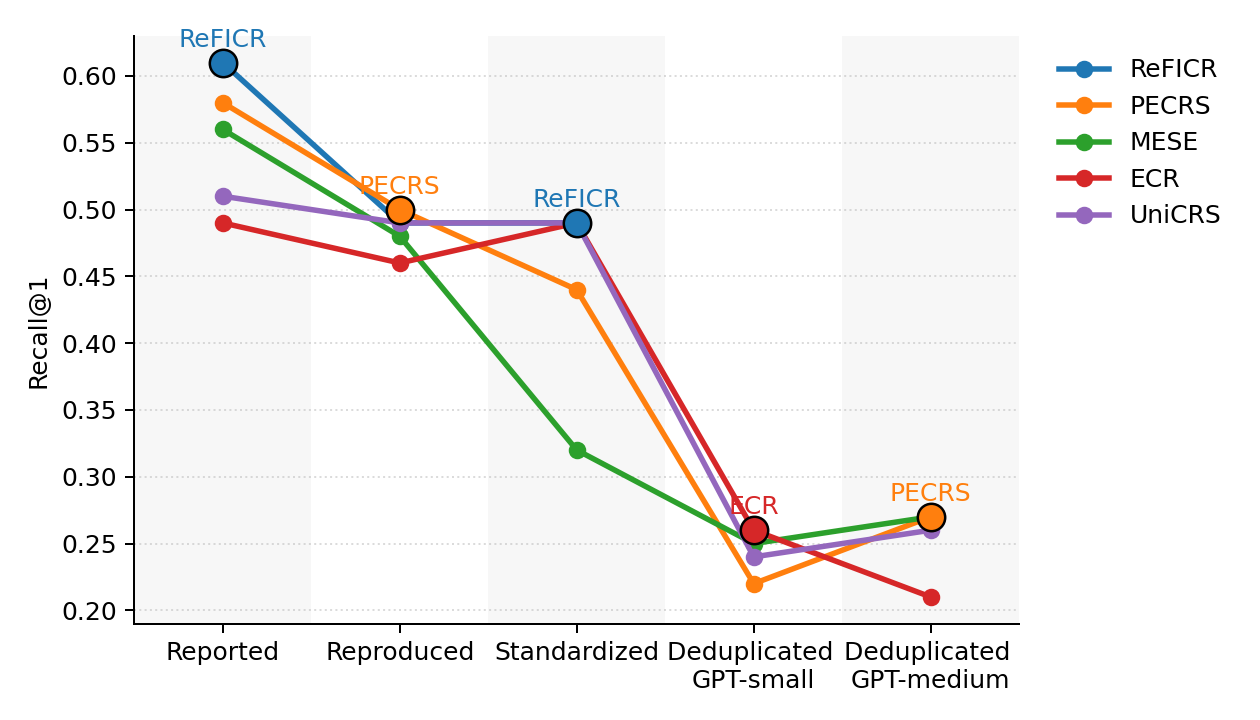}
\caption{Recall@1 changes across evaluation settings in our reproducibility study. The first column reflects the results reported in the original papers, while the remaining columns use our standardized evaluation settings across all methods. The resulting changes in ordering highlight how sensitive method comparisons are to the evaluation setup.}
    \label{fig:teaser}
\end{figure}

Conversational recommender systems (CRS) model user preferences and provide personalized item recommendations through natural language interaction. In recent years, end-to-end conversational recommender systems have become the standard approach, but they require large amounts of training data.~\citep{Jannach:2022:Artif}. Several datasets have been developed, including ReDial~\citep{Li:2018:NIPS}, INSPIRED~\citep{Hayati:2020:EMNLP}, GoRecDial~\citep{Kang:2019:EMNLP}, TG-ReDial~\citep{Zhou:2020:COLINGa}, DuRecDial~\citep{Liu:2021:EMNLP}, and LLM-REDIAL~\citep{Liang:2024:Findings} among others.

\begin{figure}
    \centering
    \includegraphics[width=0.95\linewidth]{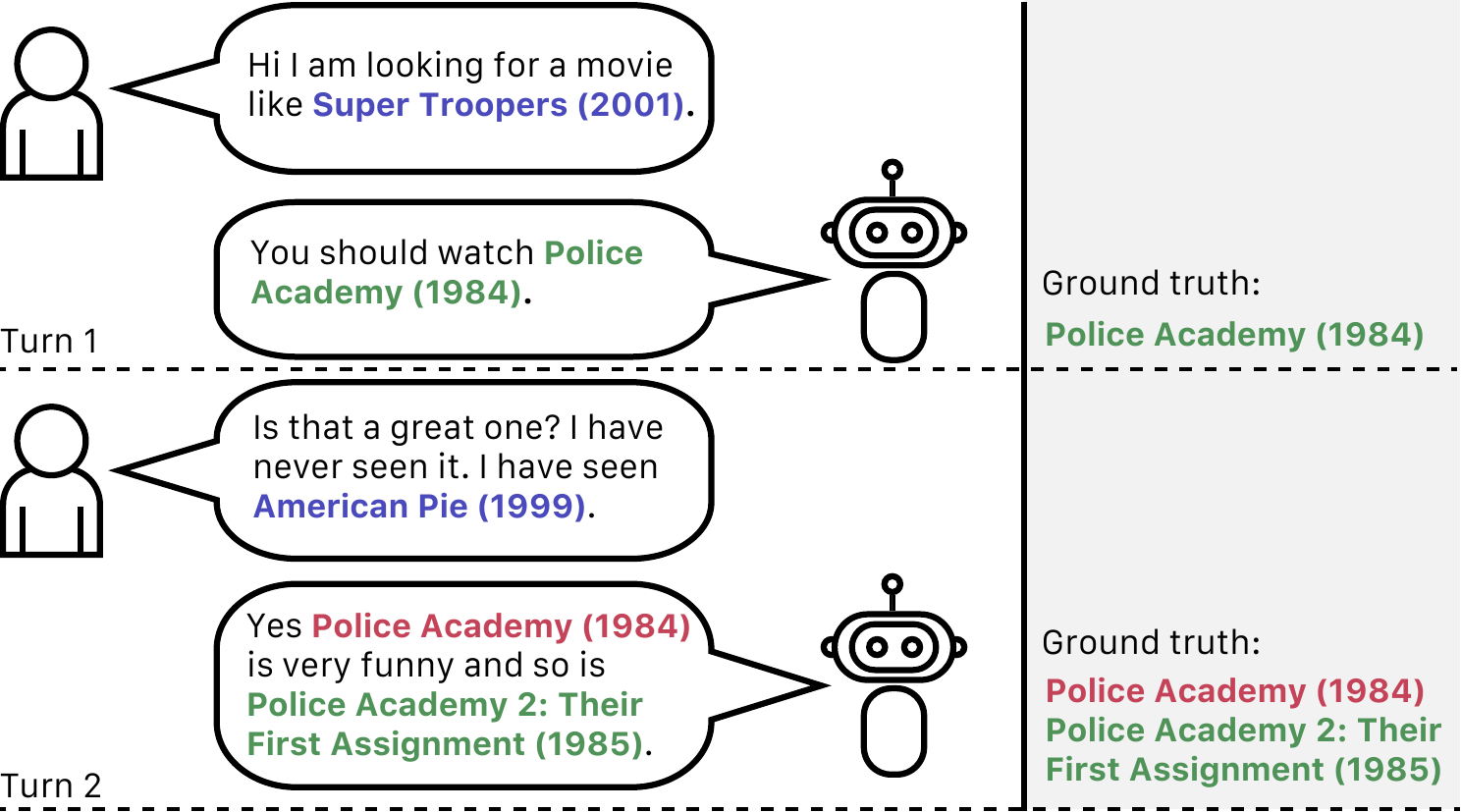}
\caption{Example dialogue from the ReDial dataset.
The dialogue is split into two evaluation instances: in Turn 1, the input is the initial seeker utterance, while in Turn 2, the input is the preceding dialogue context. Movies mentioned by the seeker are shown in {\color{darkblue}blue}, and recommendations are shown in {\color{darkgreen}green}. 
{\color{darkred}Red} highlights a recommendation that repeats an item already introduced earlier in the conversation. 
Since such repeated mentions are included in the ground truth, models can obtain artificially high scores by repeating previously mentioned movies rather than providing novel recommendations.}
    \label{fig:redial-dedupe}
\end{figure}

In this work, we focus on ReDial as it is by far the most widely used and has over 600 citations according to Google Scholar at the time of writing. 
ReDial is a large-scale human-human conversational recommendation dataset, where two crowdworkers engage in a dialogue: one acts as a seeker, stating their preferences, while the other plays the role of a recommender providing personalized recommendations throughout the conversation. The dialogues are natural language interactions and include explicit mentions of recommended items, making the dataset a standard benchmark for conversational movie recommendation (see Figure~\ref{fig:redial-dedupe} for an example dialogue).
However, the use of ReDial for training and evaluation raises several issues. Studies increasingly utilize larger and more capable large language models (LLMs) as backbones. Yet, performance gains are often attributed solely to the proposed methods, while comparisons are frequently made against baselines that use smaller or pre-LLM models. Evaluation practices also vary, as ground-truth items are often constructed by extracting mentioned movies from the conversations and cross-referenced with an external movie database, resulting in differing test sets across papers. These inconsistencies make direct comparison and reproducibility difficult. Furthermore, \citet{He:2023:CIKM} showed in their analysis that the dataset suffers from repeated recommendation issues, where the same item is recommended over multiple turns, as shown in Figure~\ref{fig:redial-dedupe}. This results in an unreasonably high performance for a naive recommender that simply recommends the last items mentioned previously in the conversation. 

More broadly, these challenges reflect concerns raised in the wider recommender systems literature about the reliability of reported progress. \citet{FerrariDacrema:2021:ACM} have shown that many claimed improvements do not come from fundamentally better modeling, but from uneven experimental conditions, such as stronger tuning effort or more modern underlying architectures for proposed methods than for baselines. At the same time, evaluation protocols often differ substantially across studies, making results difficult to compare and easy to overstate. Reproducibility is further weakened by incomplete reporting of implementation details, hyperparameter choices, and dataset construction procedures.
Additionally, \citet{Bernard:2025:SIGIR-AP} highlight two key limitations of current CRS evaluation practices: the reliance on static offline benchmarks that fail to capture interactive user behavior, and the overemphasis on recall-based accuracy metrics, which have been shown to correlate poorly with actual user satisfaction and perceived recommendation quality.


To address these challenges, we systematically re-evaluate the current landscape of ReDial-based research by controlling for three critical variables: the underlying LLM backbone, the utilization of external knowledge, and the definition of ground-truth items---specifically focusing on an evaluation setting that considers only the newly mentioned items.
We select seven influential CRS models for reproduction: KBRD~\citep{Chen:2019:EMNLP}, KGSF~\citep{Zhou:2020:KDDa}, UniCRS~\citep{Wang:2022:KDD}, ECR~\citep{Zhang:2024:RecSys}, MESE~\citep{Yang:2022:NAACL}, PECRS~\citep{Ravaut:2024:EACL}, and ReFICR~\citep{Yang:2024:RecSys}.
This selection is intentionally structured along two dimensions to capture the evolution of the field: (1) an architectural dimension, which categorizes models by how they integrate user modeling, recommendation, and response generation; and (2) a maturity dimension, which distinguishes between established, widely-cited baselines and contemporary state-of-the-art (SOTA) methods. By mapping these seven models across both architectural pipelines and temporal maturity, we can decouple performance gains derived from fundamental modeling innovations from those merely resulting from more modern backbones or differing evaluation protocols.


Our findings reveal a significant ``granularity gap'' in the reproducibility of current CRS models: while performance trends are stable at higher recall cutoffs, results at Recall@1 are highly sensitive to implementation details, with over half of the reproduced models deviating from reported scores by more than 10\%. Furthermore, our replicability analysis shows that ``repetition shortcuts''---where models suggest items already mentioned in the dialogue---account for nearly 50\% of reported accuracy. When these shortcuts are removed, even the strongest SOTA models suffer substantial performance drops. 
Our findings reveal a significant ``granularity gap'' in the reproducibility of current CRS models: while performance trends are stable at higher recall cutoffs, results at Recall@1 are highly sensitive to implementation details, with over half of the reproduced models deviating from reported scores by more than 10\%. Furthermore, our replicability analysis shows that ``repetition shortcuts''---where models suggest items already mentioned in the dialogue---account for nearly 50\% of reported accuracy. When these shortcuts are removed, even the strongest SOTA models suffer substantial performance drops. 
We also demonstrate that architectural comparisons are sensitive to the choice of LLM backbone, indicating that reported gains cannot always be attributed to architectural design alone.
When evaluation settings are standardized across methods, these effects translate into substantial performance drops and changes in the relative ranking of methods, as shown in Figure~\ref{fig:teaser}. Finally, our evaluation using novel user-centric metrics (Success Rate and RDL) confirms that high offline recall is often a poor proxy for conversational utility, as top-performing models frequently achieve accuracy at the cost of interaction efficiency.


The main contributions of this work are as follows:
\begin{itemize}
    \item We synthesize prior ReDial-based CRS approaches into a unified two-dimensional taxonomy, categorizing methods by their pipeline architectures and their temporal maturity (established vs. contemporary SOTA).
    \item We reproduce and verify the results of seven influential CRS studies across all three architectural families, establishing a consistent and comparable evaluation baseline while identifying critical sensitivities at lower recall cutoffs.
    \item We systematically re-evaluate these models under controlled conditions, isolating the impact of the underlying language model backbone, the use of external metadata, and the definition of ground-truth items.
    \item We introduce a standardized benchmarking protocol that accounts for "repetition shortcuts" and incorporates user-utility metrics (Success Rate and Reward-per-Dialogue-Length) to better reflect real-world conversational effectiveness.
    \item We revisit prior conclusions in the field, revealing how methodological inconsistencies and a reliance on static recall metrics have led to potentially overstayed or illusory performance gains.
    \item We release our complete experimental pipeline, including batch-processing extensions for legacy models and standardized data splits, to promote best practices in CRS evaluation.\footnote{\url{https://github.com/iai-group/redial-reproducibility}}.
\end{itemize}

\section{The ReDial Dataset}
\label{sec:redial}

The ReDial dataset is a large collection of human–human conversations collected using crowdworkers on Amazon Mechanical Turk. When paired, each worker was assigned a role of a seeker or a recommender. The role of the seeker is to state their preferences in the form of movie attributes (such as genre, director, or actor) or by mentioning movies they previously liked. The role of the recommender is to suggest movies that align with those preferences and to keep the dialogue flowing naturally. Each dialogue contains multiple turns in which both participants discuss movies. All movie mentions are explicitly tagged with unique identifiers. For every movie mentioned, the seeker fills a short form specifying whether they have seen it and whether they liked it, while the recommender’s form records whether the movie was suggested. A summary of the dataset statistics is shown in Table~\ref{tab:redial_stats}.

In ReDial, the dialogue can be used to evaluate two tasks: response generation and movie recommendation. Early methods trained response generation to closely match the recommender responses in the dataset. With LLMs becoming evermore capable of producing natural text, the approach has shifted to generating responses that score well using human evaluation on fluency, relevancy, and informativeness, which makes it harder and more costly to reproduce.
For this reason, in this work we focus on the recommendation task, where each recommender turn constitutes a separate evaluation instance. 
For each instance, the model receives all previous conversation turns as input context and is evaluated on its ability to recommend the ground truth movies associated with that turn. The ground truth movies are those that the human recommender mentioned in the conversation.
Figure~\ref{fig:redial-dedupe} illustrates this setup with a sample dialogue split into two evaluation instances.
The first instance uses only the context from the first seeker message and uses recommender recommendations as ground truth. The context for the second instance is the entire conversation up to the final recommender response.
Note that in the second example, the ground truth movie {\color{darkred}Police Academy (1984)}  
already appears in the conversation context, which can influence the evaluation results.  
The situation with a recommendation already appearing in the context occurs in 11.83\% of the evaluation turns in the ReDial test collection. \citet{He:2023:CIKM} found that this enables a shortcut allowing competitive recommendation results by simply recommending items already found in the conversational context. Despite this, most papers still follow the original non-realistic setup, without de-duplication, even after these concerns were acknowledged~\citep{Ravaut:2024:EACL, Yang:2024:RecSys, Zhang:2024:RecSys}.

\section{Methods}
\label{sec:method}

\begin{figure*}
    \centering
    \includegraphics[width=0.95\linewidth]{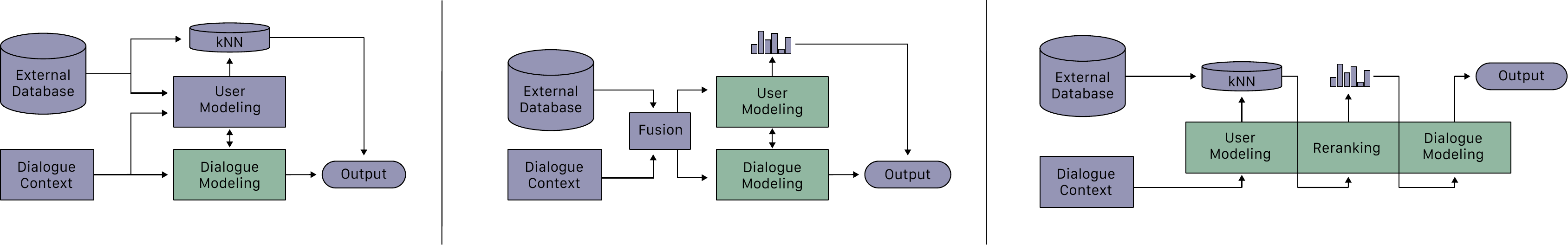}
\caption{Commonly used pipelines in modern CRSs. {\color{lightgreen} Green} signifies the components that use an LLM. 
(Left) Modular Fusion Pipelines (KBRD and KGSF) use different, disjoint components user and dialogue modeling. Outputs from the recommender are integrated with the output from the modeling components. (Middle) Shared-Backbone Pipelines (UniCRS and ECR) use the same model for recommendation and dialogue generation using a separate training procedure. 
(Right) Unified Single-Backbone Pipelines (MESE, PECRS, and ReFICR)  use truly unified approaches in which a single model is responsible for all pipeline stages and is jointly trained.}
    \label{fig:pipelines}
\end{figure*}

In this section, we describe the conversational recommender systems selected for reproduction. Prior work on ReDial spans a small number of recurring pipeline architectures (Figure~\ref{fig:pipelines}), differing mainly in how they couple user modeling, recommendation, and the use of external information. Since reproducing the full literature is infeasible, we select representative methods from each pipeline family. To ensure a comprehensive study, we further categorize our selection into two groups: (1) \emph{established methods} that serve as foundational baselines, and (2) \emph{contemporary state-of-the-art methods} that represent the current frontier of the field.

\begin{table}[t]
\centering
{\footnotesize
\begin{tabularx}{\linewidth}{lXXXX}
\toprule
\textbf{Dataset} & \textbf{\#Conv} & \textbf{\#Rec-Instances} & \textbf{\#Movie-Mentions} & \textbf{\#Unique Movies} \\
\midrule
Train & 10{,}006 & 34{,}591 & 50{,}597 & 6{,}084 \\
Test  & 1{,}342 & 4{,}198  & 6{,}736  & 1{,}936 \\
\midrule
Total & 11{,}348 & 38{,}789 & 57{,}333 & 6{,}486 \\
\bottomrule
\end{tabularx}
}
\caption{ReDial dataset statistics.}
\label{tab:redial_stats}
\end{table}

Following the taxonomy illustrated in Figure~\ref{fig:pipelines}, we organize methods into three architectural categories: (i) \textbf{Modular Fusion Pipelines}, where user modeling drives a recommender by producing a latent user representation, and the recommender output is fused with a separate dialogue model at the end~\citep{Chen:2019:EMNLP, Zhou:2020:KDDa}; (ii) \textbf{Shared-Backbone Pipelines}, where a shared backbone is used for both recommendation and response generation via a specialized, multi-stage training procedure, and the backbone is adapted primarily by learning soft prompt embeddings~\citep{Wang:2022:KDD, Zhang:2024:RecSys}; and (iii) \textbf{Unified Single-Backbone Pipelines}, where a single backbone model is responsible for user modeling, candidate ranking, and response generation, and is trained end-to-end with a multi-task objective that combines these signals into a single joint loss~\citep{Wang:2022:arXiv, Yang:2022:NAACL, Ravaut:2024:EACL, Yang:2024:RecSys}.

Table~\ref{tab:models} provides a detailed technical overview of the selected methods, categorized by their recommendation mechanisms, underlying LLM backbones, and the external knowledge sources (KG or metadata) they leverage. The evaluation coverage columns report two reproducibility constraints. \emph{Test Data} denotes the proportion of ReDial test instances that can be evaluated after applying the method-specific preprocessing and filtering steps, while \emph{Items} denotes the proportion of items from the original ReDial data collection that the method is able to recommend.

\begin{table*}[t]
\centering
\small
\caption{Overview of the selected established (\established) and contemporary state-of-the-art (\sota) methods for the reproducibility study, focusing on evaluation settings.} 
\label{tab:models}
\begin{tabular*}{\linewidth}{@{\extracolsep{\fill}}clccllcccc}
\toprule
\multirow{2}{*}{Pipeline} 
& \multirow{2}{*}{Method} 
& \multicolumn{2}{c}{Rec Mechanism} 
& \multicolumn{2}{c}{Backbone} 
& \multicolumn{2}{c}{External Data} 
& \multicolumn{2}{c}{Evaluation Coverage}  \\
\cmidrule(lr){3-4}
\cmidrule(lr){5-6}
\cmidrule(lr){7-8}
\cmidrule(lr){9-10}
 & & NN & Seq2Seq 
 & Primary Model & Secondary Models
 & KG & Metadata 
 & Test Data 
 & Items  
 \\
\midrule
\multirow{2}{*}{(i)} & \established~KBRD~\citep{Chen:2019:EMNLP}    
& \cmark  &  & Transformer & RGCN & \cmark &  
& 82.6\% & 87.3\% \\
& \established~KGSF~\citep{Zhou:2020:KDDa}       
& \cmark  &  & Transformer & RGCN,  & \cmark & 
& 82.6\%  & 87.3\% \\
\midrule
\multirow{2}{*}{(ii)} & \established~UniCRS~\citep{Wang:2022:KDD} 
&  & \cmark & DialoGPT-small & R-GCN, RoBERTa & \cmark & 
& 93.5\% & 96.8\% \\
& \sota~ECR~\citep{Zhang:2024:RecSys}     
&   & \cmark & DialoGPT-small & R-GCN, RoBERTa & \cmark & 
& 93.5\% & 96.8\% \\
\midrule
\multirow{3}{*}{(iii)} & \sota~MESE~\citep{Yang:2022:NAACL}      
& \cmark  & \cmark & GPT2-small & DistilBERT &  & \cmark
& 78.3\% &  83.7\% \\
& \sota~PECRS~\citep{Ravaut:2024:EACL}   
& \cmark & \cmark  & GPT2-medium & --- &  & \cmark
& 95.6\% & 95.2\% \\
& \sota~ReFICR~\citep{Yang:2024:RecSys}   
& \cmark  & \cmark & GritLM-7B (Mistral) & --- &  & \cmark
& 94.2\% & 93.3\%  
\\
\bottomrule
\end{tabular*}
\end{table*}

\subsection{Modular Fusion Pipelines}
\boldheading{KBRD}~\citep{Chen:2019:EMNLP} is an entity-centric conversational recommender that extracts mentioned entities from the dialogue, links them to items in a knowledge graph (KG) (DBPedia), and ranks candidate items by similarity matching in the KG embedding space. The authors attribute its gains primarily to explicit entity tracking and reasoning over the knowledge graph. They argue that KG paths capture user intent more reliably than surface text alone.

\boldheading{KGSF}~\citep{Zhou:2020:KDDa} extends KBRD by introducing semantic fusion mechanisms that combine conversational context representations (ConceptNet\footnote{https://conceptnet.io/}) with KG-based item embeddings (DBpedia). The paper claims improved performance stems from better alignment between textual dialogue signals and structured KG semantics.

\subsection{Shared-Backbone Pipelines}
\boldheading{UniCRS}~\citep{Wang:2022:KDD} reformulates conversational recommendation as a sequence-to-sequence generation task, where recommendations are produced directly from dialogue context augmented with knowledge-enhanced prompts. The authors argue that unifying recommendation and response generation allows the model to better leverage contextual cues.

\boldheading{ECR}~\citep{Zhang:2024:RecSys} models conversational recommendation as an empathetic response generation task, conditioning recommendations on inferred user emotions and preferences expressed in dialogue. It is an extension of UniCRS, and the paper claims that emotionally aligned recommendations improve ranking performance.

\subsection{Unified Single-Backbone Pipelines}
\boldheading{MESE}~\citep{Yang:2022:NAACL} incorporates explicit item metadata into a neural recommendation framework, using context-aware encoding to align dialogue representations with item attributes. The dialogue and recommendation components share the same model, which is trained with a joint loss. Performance gains are attributed to richer semantic grounding from metadata.

\boldheading{PECRS}~\citep{Ravaut:2024:EACL} is an extension of MESE, and frames conversational recommendation as a parameter-efficient language modeling task, adapting pretrained models with lightweight tuning modules (prompt tuning and LoRA~\cite{Dettmers:2023:arXiv}). The authors claim that strong results arise from effective reuse of pretrained knowledge rather than task-specific modeling.

\boldheading{ReFICR}~\citep{Yang:2024:RecSys} treats conversational recommendation as a retrieval problem, using dense representations of dialogue context and item metadata for similarity-based ranking. Trained with a contrastive loss objective, the paper attributes its superior performance to improved retrieval alignment.

\section{Reproducibility}
\label{sec:rep}

In this section, we report the main results of our reproducibility study. We compare the performance metrics reported in the original papers with the results obtained by using the resources and models accompanying those papers. 

\subsection{Protocol}

We follow a standardized protocol across all methods: our goal is to match each original paper’s reported setting as closely as possible, while minimizing implementation variance. For each method, we first verify whether the authors provide a pretrained checkpoint. When available, we use the released checkpoint as the primary artifact (e.g., ECR, ReFICR). For KBRD and KGSF, we use the widely adopted benchmark implementation in CRSLab~\citep{Zhou:2021:IJCNLP} to mitigate reimplementation differences and to align with common community practice for these models. For all remaining methods lacking an official released checkpoint (UniCRS, MESE, PECRS), we trained the models using the authors’ publicly available codebase, following the reported training procedure and the default repository configuration when details were unspecified. To ensure reproducibility, we use the authors’ released preprocessed data when provided (KBRD, KGSF, UniCRS, ECR). Otherwise, we reproduced the required inputs by running the preprocessing scripts from the corresponding repository (MESE, PECRS, ReFICR). This includes any repository-specific formatting needed to construct training, validation, and test instances. Most methods follow a similar preprocessing pipeline: merging consecutive turns from the same speaker into a single utterance, discarding recommendation turns with no previous conversational context, and filtering out movies that cannot be linked to the external movie database required by the method.

\subsection{Execution}

During the execution phase, we found that reproducing results required non-trivial engineering effort despite the public availability of code. Two out of seven repositories (MESE and UniCRS) used default configurations that deviated from the paper (model hyperparameters), requiring manual alignment to the reported setup. Three repositories (MESE, PECRS, and UniCRS) had minor but blocking issues that needed fixes before running, most commonly hardcoded local paths, missing file references, and brittle scripts that assumed a specific directory structure or environment. Only CRSLab provided end-to-end documentation sufficient to reproduce results without extensive code inspection. Three repositories (UniCRS, ECR, and ReFICR) offered only minimal instructions, with critical details (preprocessing choices, evaluation protocol, and expected outputs) either missing or implicit, while two repositories (MESE and PECRS) only pointed to the main file without additional guidance. We also observed runtime as a practical reproducibility constraint: two methods (MESE and PECRS) implement sequential (non-batched) training or inference, which dramatically increases training and inference time and makes hyperparameter search or ablations impractical on typical hardware.

\subsection{Results}

\begin{table}[t]
\centering
\small
\caption{Main reproducibility results, comparing the numbers reported in the original paper with our reproduced numbers. 
Colors indicate the relative performance difference between the original paper and our implementation: \textbf{\textcolor{green!60!black}{green}} ($\le 5\%$), \textbf{\textcolor{orange}{orange}} ($> 5\%$ and $\le 10\%$), and \textbf{\textcolor{red}{red}} ($> 10\%$).
}
\label{tab:main_results}
\begin{tabular*}{\linewidth}{@{\extracolsep{\fill}}lcccccc}
\toprule
\multirow{2}{*}{Method} 
& \multicolumn{3}{c}{Original Paper} 
& \multicolumn{3}{c}{Our Implementation} \\
\cmidrule(lr){2-4}
\cmidrule(lr){5-7}
& R@1 & R@10 & R@50 
& R@1 & R@10 & R@50 \\
\midrule
KBRD~\citep{Chen:2019:EMNLP}    
& 0.030 & 0.163 & 0.338 
& \textcolor{red}{0.036} \textcolor{green!60!black}{$\uparrow$} & \textcolor{orange}{0.176} \textcolor{green!60!black}{$\uparrow$} & \textcolor{green!60!black}{0.334} \textcolor{red}{$\downarrow$} \\
KGSF~\citep{Zhou:2020:KDDa}     
& 0.039 & 0.183 & 0.378 
& \textcolor{red}{0.034} \textcolor{red}{$\downarrow$} & \textcolor{green!60!black}{0.179} \textcolor{red}{$\downarrow$} & \textcolor{green!60!black}{0.365} \textcolor{red}{$\downarrow$} \\
UniCRS~\citep{Wang:2022:KDD}    
& 0.051 & 0.224 & 0.428
& \textcolor{green!60!black}{0.049} \textcolor{red}{$\downarrow$} & \textcolor{green!60!black}{0.213} \textcolor{red}{$\downarrow$} & \textcolor{green!60!black}{0.421} \textcolor{red}{$\downarrow$} \\
ECR~\citep{Zhang:2024:RecSys}   
& 0.049 & 0.220 & 0.428 
& \textcolor{orange}{0.046} \textcolor{red}{$\downarrow$} & \textcolor{green!60!black}{0.217} \textcolor{red}{$\downarrow$} & \textcolor{green!60!black}{0.426} \textcolor{red}{$\downarrow$} \\
MESE~\citep{Yang:2022:NAACL}   
& 0.056 & 0.256 & 0.455  
& \textcolor{red}{0.048} \textcolor{red}{$\downarrow$}  & \textcolor{orange}{0.243} \textcolor{red}{$\downarrow$} & \textcolor{green!60!black}{0.452} \textcolor{red}{$\downarrow$}   \\
PECRS~\citep{Ravaut:2024:EACL}  
& 0.058 & 0.225 & 0.416
& \textcolor{red}{0.050} \textcolor{red}{$\downarrow$} & \textcolor{orange}{0.211} \textcolor{red}{$\downarrow$} & \textcolor{green!60!black}{0.401} \textcolor{red}{$\downarrow$} \\
ReFICR~\citep{Yang:2024:RecSys} 
& 0.061 & 0.305 & 0.532 
& \textcolor{red}{0.049} \textcolor{red}{$\downarrow$} & \textcolor{orange}{0.280} \textcolor{red}{$\downarrow$} & \textcolor{green!60!black}{0.522} \textcolor{red}{$\downarrow$}      \\
\bottomrule
\end{tabular*}
\end{table}

Our main reproducibility results, comparing recall reported in the original papers with the results obtained using the authors' released code, data, and (when available) checkpoints, are shown in Table~\ref{tab:main_results}.  Overall, we obtained very similar performance for all methods for higher recall cutoffs (R@10 and R@50) and within 5\% margin for only one out of seven methods on R@1: for UniCRS, all three recall cutoffs differ by at most 4.9\% relative to the original reports; ECR is similarly close on R@10 and R@50 (with up to 1.4\% difference), with a slightly larger deviation on R@1 (6.1\%). The largest discrepancies are observed for the two earliest KG-based methods. For KBRD, the relative difference reaches 20.0\% on R@1, while R@10 and R@50 remain closer (8.0\% and 1.2\%, respectively). For KGSF, the maximum relative difference is 12.8\% on R@1, with smaller gaps on R@10 and R@50 (2.2\% and 3.4\%). We observed that R@10 and R@50 generally increased steadily across training epochs for all methods, whereas R@1 showed larger oscillations. This makes the choice of training run or checkpoint particularly consequential when reporting R@1. While we cannot verify the run-selection procedure used in the original papers, selecting the run with the strongest test set results rather than reporting a single randomly initialized run could lead to noticeable differences between the reported and reproduced R@1 results.

\section{Replicability}
\label{sec:replicability}

In this section, we study replicability, i.e., whether prior conclusions remain consistent under standardized, method-independent evaluation settings that are explicitly defined and comparable across methods. To make this systematic large-scale evaluation feasible, we first addressed a significant technical bottleneck found in several original codebases: the lack of efficient inference. To speed up evaluation, we implemented batch processing for all methods that were missing it, reducing the evaluation time from days to hours (Section~\ref{sec:repl:exec}). 

We structure this analysis around two key aspects: standardization and generalization.
By standardization (Section~\ref{sec:standard}), we refer to the enforcement of uniform experimental conditions—including identical data splits, preprocessing pipelines, and evaluation protocols—to eliminate ``hidden'' advantages caused by idiosyncratic turn-filtering or differing ground-truth definitions.
By generalization (Section~\ref{sec:gen}), we investigate the stability of architectural claims when subjected to change. This involves testing whether observed performance gains are truly inherent to a specific architecture or are merely artifacts of a particular LLM backbone, and whether these improvements hold when measured against conversational utility metrics that go beyond simple recall.

\subsection{Execution}
\label{sec:repl:exec}

To produce a standardized evaluation environment, we implemented a rigorous preprocessing pipeline on the original ReDial dataset. First, we flattened the human-human conversations into individual turns with their associated dialogue context. During this process, we merged all consecutive utterances from the same speaker into single turns to maintain logical flow. We then filtered the dataset to remove turns that lacked either a recommendation or sufficient prior dialogue context (e.g., cases where the recommender suggests an item in the very first turn).

For each evaluated method, we mapped movie mentions to their respective external sources, such as DBPedia or specific metadata databases. If a mention was missing from a method's specific database, we removed it from the conversational context to prevent model errors, while crucially retaining it in the ground truth with a unique negative identifier (e.g., -101). This ensures that recall calculations remain standardized and penalized across all methods equally, regardless of their specific item coverage. Finally, to facilitate our analysis of the ``repetition shortcut,'' we generated a deduplicated version of the test set by adding a filter that removes any ground-truth items that had already appeared in the preceding conversational context.

To speed up evaluation we implemented batch processing for the methods that were missing it (MESE and PECRS). This reduced inference time on these methods by a factor of 20.

\subsection{Standardization}
\label{sec:standard}

\begin{table}[t]
\centering
\small
\caption{Reproduced results on standardized dataset with and without de-duplication. As a reference we include a naive method that recommends context items in reversed order of mentions.}
\label{tab:dedup}
\begin{tabular*}{\linewidth}{@{\extracolsep{\fill}}lcccccc}
\toprule
\multirow{2}{*}{Method} 
& \multicolumn{3}{c}{Without Deduplication} 
& \multicolumn{3}{c}{With Deduplication} \\
\cmidrule(lr){2-4}
\cmidrule(lr){5-7}
& R@1 & R@10 & R@50 
& R@1 & R@10 & R@50 \\
\midrule
KBRD~\citep{Chen:2019:EMNLP}    
& 0.033 & 0.177 & 0.332 
& 0.017 & 0.148 & 0.300 \\
KGSF~\citep{Zhou:2020:KDDa}     
& 0.032 & 0.177 & 0.369
& 0.016 & 0.147 & 0.327 \\
UniCRS~\citep{Wang:2022:KDD}    
& 0.049 & 0.213 &  0.421
& 0.024 & 0.180 & 0.335 \\
ECR~\citep{Zhang:2024:RecSys}   
& 0.049 & 0.220 & 0.428 
& 0.026 & 0.181 & 0.337 \\
MESE~\citep{Yang:2022:NAACL}   
& 0.032  &  0.178 & 0.381
& 0.025 & 0.159 &  0.361 \\ 
PECRS~\citep{Ravaut:2024:EACL} 
& 0.044 & 0.189 & 0.383
& 0.022 & 0.143 & 0.341 \\
ReFICR~\citep{Yang:2024:RecSys} 
& 0.049 & 0.270 & 0.507
& 0.018 & 0.213 & 0.465 \\
\midrule
Naive & 0.043 & 0.090 & 0.090
& 0.000 & 0.000 & 0.000 \\
\bottomrule
\end{tabular*}
\end{table}

The findings from Table~\ref{tab:dedup} reveal significant insights into the nature of reported performance on the ReDial dataset, specifically regarding the ``repetition bias'' and its impact on the perceived state-of-the-art.

The most striking result is the precipitous drop in performance across all methods when repeated recommendations are removed. At the R@1 cutoff, performance for most models effectively halves upon deduplication. For instance, UniCRS and ECR drop from 0.049 to 0.024 and 0.026, respectively. This suggests that a substantial portion of the models' ``accuracy'' is derived from simply predicting movies that the seeker or recommender has already mentioned in the preceding dialogue turns.

The influence of this shortcut is best illustrated by the Naive baseline, which simply recommends context items in reverse order of mention. In the Without Deduplication setting, this simple heuristic achieves an R@1 of 0.043, outperforming several complex, trained models, such as KBRD, KGSF, and MESE (all $\approx 0.032$). Even compared to the best-performing models (0.049), the naive baseline captures nearly 88\% of the achieved R@1. Once deduplicated, the naive baseline naturally falls to 0.000, revealing the extent to which standard evaluation metrics are inflated by ``low-effort'' correct predictions.

We further observe that the gap between the strongest and weakest methods does not narrow when evaluating true novelty. Without deduplication, the top-performing R@1 methods (UniCRS, ECR, and ReFICR) outperform the weakest methods (KGSF and MESE) by approximately 53\% relative on R@1 (0.049 vs. 0.032). With deduplication, the strongest method at R@1 becomes ECR, which outperforms the weakest method, KGSF, by approximately 62.5\% relative (0.026 vs. 0.016). ReFICR suffers a substantial relative drop of 63.3\% in R@1 performance, falling from 0.049 to 0.018. The relative ranking of methods also undergoes a notable shift. In the standard evaluation without deduplication, ReFICR is tied for the best R@1 performance and is the clear leader at R@10 (0.270) and R@50 (0.507). While it remains the strongest at higher recalls after deduplication, with R@10 = 0.213 and R@50 = 0.465, its R@1 ranking collapses. In the deduplicated setting, ECR (0.026) and MESE (0.025) emerge as the top-performing models at R@1, while ReFICR falls to 0.018, only slightly above KBRD (0.017), an established method from 2019. This indicates that while contemporary SOTA models remain better at retrieving a broader set of relevant novel items at R@10 and R@50, they are significantly less effective at identifying the specific next new item once repeated context items are removed.

\subsubsection{Architectural Stability}

\begin{figure}
    \centering
    \includegraphics[width=\linewidth]{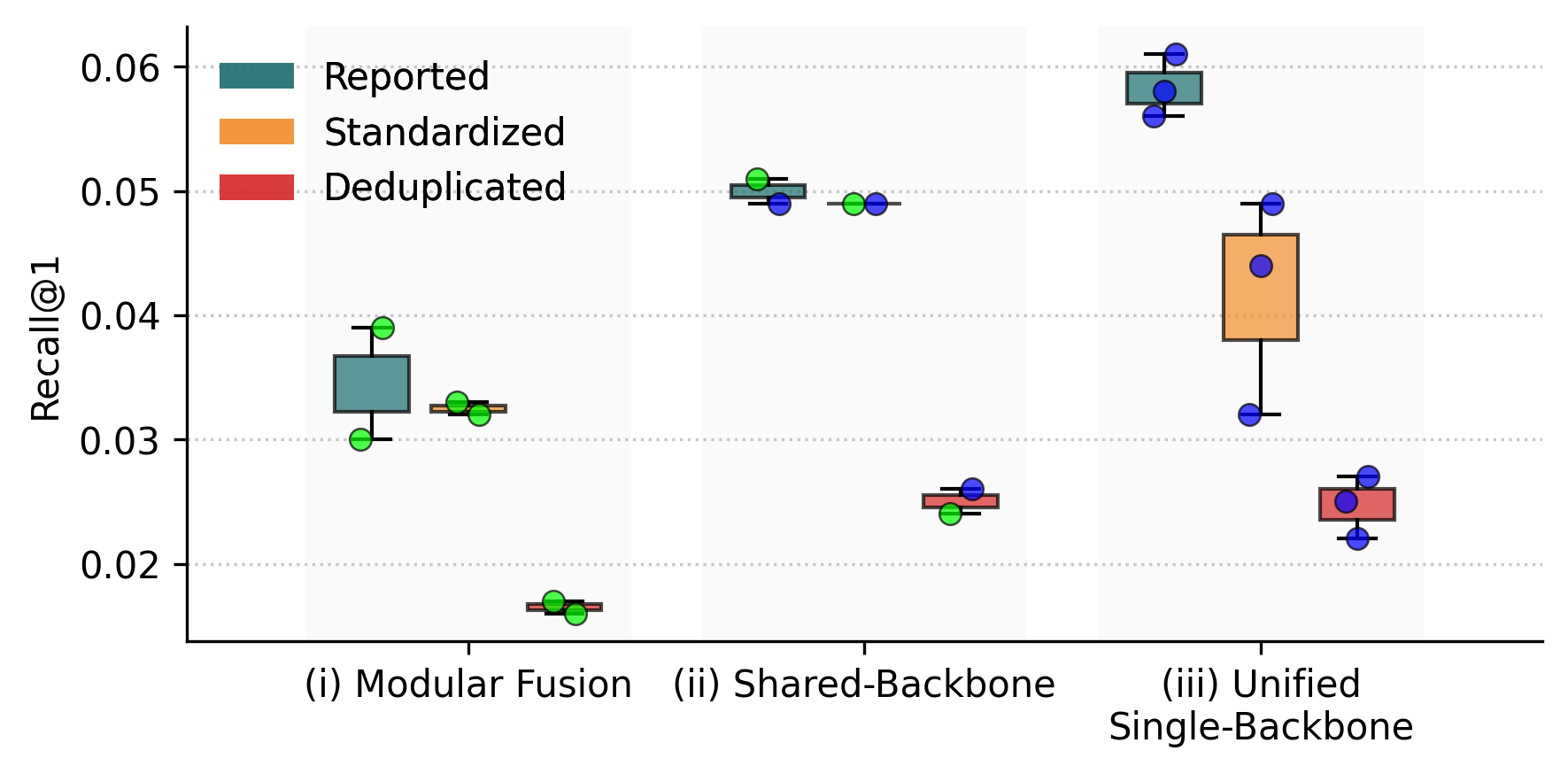}
\caption{Recall@1 under reported, standardized, and deduplicated evaluation settings, grouped by CRS architecture family. Points indicate established (\established) and contemporary state-of-the-art (\sota) methods.}
    \label{fig:boxplot_differences}
\end{figure}

Figure~\ref{fig:boxplot_differences} shows how evaluation changes affect methods across both architectural families and maturity groups. Performance decreases across all architectures after standardization and deduplication, with the magnitude of the decrease varying by architecture.  Groups (i) and (ii) show relatively small changes from reported to standardized results, whereas group (iii) shifts more substantially. The strongest reported scores are concentrated in group (iii); however, this group also shows the greatest sensitivity to the evaluation setting.

\subsection{Generalization}
\label{sec:gen}

In this section, we evaluate the generalizability of the proposed CRS architectures across two critical dimensions: backbone invariance and metric robustness. Our primary objective is to disentangle performance gains: we seek to determine if reported improvements stem from superior architectural design choices or are simply a byproduct of utilizing more powerful LLM backbones.

To test this, we establish a standardized experimental framework where each CRS architecture is decoupled from its original backbone and re-trained using a common set of base models. 
We limit this analysis to UniCRS, ECR, MESE, and PECRS, since these methods all rely on the GPT2-family backbone group and can therefore be compared under a controlled backbone replacement setting.
We select three specific backbones for this cross-comparison: GPT2-small, GPT2-medium, and DialoGPT-small. These models have been chosen because GPT2 and DialoGPT are the most common backbone families among the evaluated methods. DialoGPT is architecturally compatible with GPT2 while being further trained on dialogue data, making it straightforward to substitute in most implementations. By comparing architectures across these models, we can observe whether the relative performance hierarchy remains stable as the underlying language modeling capability changes. 
We exclude KBRD and KGSF because they do not use pretrained language-model backbones. We further exclude ReFICR because it is optimized around GritLM-7B, a much larger instruction-tuned generative-representational model. Forcing these methods into the same GPT2-based setup would therefore conflate architectural differences with incompatible modeling and training assumptions. 
One caveat is that all three backbone models were released after ReDial became available, creating a potential risk of data contamination. This, however, should not affect the relative comparisons within each backbone condition because the same backbones are applied uniformly across all methods.

Furthermore, we investigate whether the findings reported in prior literature generalize to alternative evaluation measures. Traditional recall-based metrics have significant shortcomings in the CRS domain, most notably their failure to account for the interactive, conversational nature of the task. To address this, we re-evaluate the models using a suite of recently proposed metrics—specifically those developed to capture dialogue quality and user intent—which have been shown to correlate more strongly with actual user satisfaction than offline recall~\citep{Bernard:2025:SIGIR-AP}.

\subsubsection{LLM Backbone}

\begin{table}[t]
\centering
\small
\caption{Cross-backbone generalizability results on the standardized deduplicated test collection. Performance is reported for the four LLM-based methods using three standardized transformer backbones to isolate architectural gains from model capacity.}
\label{tab:rep}
\begin{tabular*}{\linewidth}{@{\extracolsep{\fill}}lcccccc}
\toprule
\multirow{2}{*}{Method} 
& \multicolumn{2}{c}{GPT2-small} 
& \multicolumn{2}{c}{GPT2-medium} 
& \multicolumn{2}{c}{DialoGPT-small} \\
\cmidrule(lr){2-3}
\cmidrule(lr){4-5}
\cmidrule(lr){6-7}
& R@1 & R@50 & R@1 
& R@50 & R@1 & R@50 \\
\midrule
UniCRS~\citep{Wang:2022:KDD}    
& 0.021 & 0.315
& 0.026 & 0.321
& 0.024  & 0.335\\
ECR~\citep{Zhang:2024:RecSys}   
& 0.019 & 0.315
& 0.021 & 0.330
& 0.026  & 0.337 \\
MESE~\citep{Yang:2022:NAACL}   
& 0.025 & 0.361
& 0.027 & 0.371
& 0.016 & 0.312   \\ 
PECRS~\citep{Ravaut:2024:EACL} 
& 0.022 & 0.341
& 0.027 & 0.362
& 0.017 & 0.345 \\
\bottomrule
\end{tabular*}
\end{table}

Table~\ref{tab:rep} presents the performance of the four LLM-based CRS architectures when controlled for the underlying backbone on the deduplicated test set. By standardizing the backbone, we can isolate the contribution of the proposed CRS frameworks from the generative capabilities of the base models.

\paragraph{Backbone Scaling and Sensitivity}
We observe that increasing the model capacity from GPT2-small to GPT2-medium yields consistent, albeit modest, improvements across all architectures. For example, UniCRS and PECRS see R@1 gains of 23.8\% and 22.7\% relative, respectively, when moving to the larger backbone. This suggests that the complex reasoning required for conversational recommendation---even after removing repetition shortcuts---is positively correlated with the number of parameters in the underlying LLM.

\paragraph{Dialogue-specific Pretraining}
The results for DialoGPT-small reveal a divergence in architectural robustness. For Shared-Backbone Pipelines (UniCRS, ECR), DialoGPT-small consistently outperforms GPT2-small at R@50, confirming that dialogue-centric pretraining provides a better initialization for response-driven recommendation. However, for Unified Single-Backbone Pipelines (MESE and PECRS), we observe a significant performance degradation when moving to DialoGPT. MESE drops from 0.361 to 0.312 in R@50, while PECRS improves slightly from 0.341 to 0.345 in R@50, but drops from 0.022 to 0.017 in R@1. This suggests that the training objectives used for DialoGPT may interfere with the multi-task joint loss optimization used in unified architectures, whereas generative-heavy pipelines (ECR) are more resilient to this change.

\paragraph{Architectural Ranking Stability}
The relative ranking of the methods remains partly stable across backbones. MESE and PECRS consistently remain among the strongest methods, although their relative ordering varies depending on the backbone and metric. This suggests that the architectural innovations proposed in these papers---such as metadata integration and parameter-efficient tuning---do provide genuine modeling advantages that are invariant to the choice of the underlying LLM.

\subsubsection{Conversational Metrics}

Standard offline evaluation metrics like Recall@$N$ often focus on narrow system aspects—specifically recommendation accuracy—while failing to capture overall user utility. As argued by \citet{Bernard:2025:SIGIR-AP}, the prevalent use of large $N$ values ($N > 10$) is poorly suited for conversational settings where users expect concise, targeted suggestions rather than a deluge of items that may overwhelm them. Furthermore, recent user studies have found a negligible correlation between Recall@10 and self-reported user satisfaction, suggesting that these traditional metrics are increasingly detached from real-world utility. To bridge this gap, we re-evaluate the selected models using a reward/cost framework that lift the assumption of a single target item and explicitly accounts for the interactive nature of the task. We report on two of the best performing conversational metrics proposed in \citep{Bernard:2025:SIGIR-AP}:
\begin{itemize}
    \item \textbf{Success Rate (SR):} The average proportion of dialogues where the system successfully recommends at least one relevant item.
    \item \textbf{Reward-per-Dialogue-Length (RDL):} A utility-based metric that assigns a reward of 1.0 for a liked recommendation and 0.5 for a relevant recommendation already seen by the user, normalized by the total number of turns in the dialogue to penalize excessive interaction effort.
\end{itemize}

\begin{table}[t]
\centering
\small
\caption{User-centric utility metrics on the standardized dataset. Success Rate (SR) and Reward-per-Dialogue-Length (RDL) are reported alongside R@1 to compare recommendation accuracy against interaction efficiency, with and without de-duplication.}
\label{tab:standard_sr_rdl}
\begin{tabular*}{\linewidth}{@{\extracolsep{\fill}}lcccccc}
\toprule
\multirow{2}{*}{Method}
& \multicolumn{3}{c}{Without Deduplication}
& \multicolumn{3}{c}{With Deduplication} \\
\cmidrule(lr){2-4}
\cmidrule(lr){5-7}
& R@1 & SR & RDL  & R@1 & SR & RDL \\
\midrule
KBRD~\citep{Chen:2019:EMNLP}     & 0.033 & 0.103 & 0.049 & 0.017 & 0.046 & 0.036 \\
KGSF~\citep{Zhou:2020:KDDa}      & 0.032 & 0.093 & 0.048 & 0.016 & 0.036 & 0.036 \\
UniCRS~\citep{Wang:2022:KDD}     & 0.049 & 0.101 & 0.041 & 0.024 & 0.048 & 0.030 \\
ECR~\citep{Zhang:2024:RecSys}    & 0.049 & 0.104 & 0.041 & 0.026 & 0.063 & 0.033 \\
MESE~\citep{Yang:2022:NAACL}     & 0.032 & 0.113 & 0.051 & 0.025 & 0.051 & 0.043 \\
PECRS~\citep{Ravaut:2024:EACL}   & 0.044 & 0.121 & 0.053 & 0.022 & 0.052 & 0.047 \\
ReFICR~\citep{Yang:2024:RecSys}  & 0.049 & 0.151 & 0.074 & 0.018 & 0.053 & 0.053 \\
\midrule
Naive & 0.043 & 0.100 & 0.009 & 0 & 0 & 0 \\
\bottomrule
\end{tabular*}
\end{table}

\subsubsection{Results}
The results in Table~\ref{tab:standard_sr_rdl} reveal that high recall does not always translate to high conversational utility. In the Without Deduplication setting, UniCRS, ECR, and ReFICR achieve the highest R@1 (0.049), but their utility scores differ substantially. UniCRS and ECR obtain relatively low RDL scores among trained models (0.041 and 0.041), whereas ReFICR achieves the highest RDL (0.074). This indicates that identical first-rank accuracy can still correspond to very different recommendation behavior across the full dialogue. In contrast, PECRS achieves a higher RDL (0.053), despite having a lower R@1 (0.044), suggesting that although it less often matches the exact movie recommended in the original ReDial dialogue at rank 1, its recommendations more often align with movies the user marked as liked or already seen.

Under deduplication, the metrics reveal a substantial reduction in utility for several models, but the effect is not uniform. For UniCRS and ECR, RDL drops to 0.030 and 0.033, respectively. ReFICR no longer remains the strongest model in terms of R@1 after deduplication, falling to 0.018, while ECR (0.026) and MESE (0.025) achieve the highest R@1 scores. However, ReFICR retains the highest RDL (0.053) and the second-highest SR (0.053), behind ECR (0.063). PECRS also remains strong in terms of utility, achieving SR = 0.052 and RDL = 0.047. These findings confirm the hypothesis that traditional recall metrics can be insensitive to user-centric utility, since models with similar or lower R@1 can differ substantially in accumulated recommendation reward under the same ReDial dialogue structure.

\section{Conclusion}
\label{sec:con}

In this work, we have addressed the growing concerns regarding reproducibility and benchmarking consistency in conversational recommender systems. By focusing on the widely utilized ReDial dataset, we synthesized recent research into a unified taxonomy defined by two key dimensions---architectural pipeline and temporal maturity---and selected seven representative models for reproduction. 

Our reproducibility study reveals a significant "granularity gap" in the reliability of reported results. While absolute performance trends remain stable at higher recall cutoffs---where all seven methods were successfully reproduced within a 5\% relative difference at Recall@50---reproducibility at the top of the ranked list is notably problematic. At the Recall@1 cutoff, five out of the seven methods fell outside the 10\% relative difference threshold, suggesting that the fine-grained ranking capabilities of these systems are highly sensitive to local environment configurations, random seeds, or undocumented preprocessing choices.

Beyond the immediate results of reproduction, our replicability study reveals the extent to which current state-of-the-art claims are contingent upon specific evaluation protocols and model choices. By standardizing the experimental conditions—including identical data splits and the removal of "repetition shortcuts"—we found that model performance is significantly lower than previously reported. Specifically, when evaluating true novelty by deduplicating ground-truth items from the conversational context, Recall@1 scores for most models dropped by nearly 50\%. This suggests that a substantial portion of reported progress in the field is driven by models exploiting the dataset's tendency to repeat previously mentioned items, a shortcut effectively captured by our naive baseline. Even for ReFICR, which remains the strongest model at Recall@10 and Recall@50 after deduplication and achieves the highest RDL, its Recall@1 falls behind several smaller models, despite being newer and substantially larger.

Our generalization analysis further clarifies the drivers of this performance. We found that architectural improvements are often secondary to the choice of the underlying LLM backbone; transitioning from GPT2-small to GPT2-medium provided consistent gains that frequently outweighed the benefits of specific architectural innovations. Furthermore, following the framework of \citet{Bernard:2025:SIGIR-AP}, our re-evaluation using conversational utility metrics (SR and RDL) demonstrates that high offline recall does not necessarily correlate with user-centric success. Models that achieve top-tier recall often do so at the cost of dialogue efficiency, while unified architectures like PECRS demonstrate superior utility by providing relevant suggestions in fewer interaction turns. 

These findings strongly reinforce long-standing concerns in the wider information retrieval and recommender systems literature regarding the reliability of evaluation practices and the potential for ``illusory'' progress. Our results suggest that the field is currently vulnerable to architectural gains being confounded by backbone strength and evaluation metrics that are detached from user utility. By systematically controlling for these variables and establishing a rigorous, open-source benchmarking pipeline, our study makes a significant step towards standardizing CRS evaluation. We argue that promoting these best practices—specifically novelty-focused deduplication and the adoption of utility-based conversational metrics—is essential for ensuring that the next generation of conversational assistants delivers genuine value to users.

\begin{acks}
    An unrestricted gift from Google partially supported this research.
\end{acks}

\bibliographystyle{ACM-Reference-Format}
\balance
\bibliography{sigir2026-crs-repro}

\end{document}